\documentclass[aps,prl,twocolumn,floatfix,english,showpacs,10pt,superscriptaddress]{revtex4-2}%
\usepackage{graphicx}
\usepackage{booktabs}
\usepackage{amsmath}
\usepackage{physics}
\usepackage{amssymb}
\usepackage{mathrsfs}
\usepackage{colordvi}
\usepackage{verbatim}
\usepackage{xcolor}
\usepackage{multirow}
\usepackage{mathrsfs}
\usepackage{epsfig}
\usepackage{lipsum}
\usepackage{enumitem}
\usepackage{amsfonts}

\usepackage[unicode=true, breaklinks=false, pdfborder={0 0 1}, backref=false,
colorlinks=true, linkcolor=blue, urlcolor=blue, citecolor=blue]{hyperref}%
\providecommand{\U}[1]{\protect\rule{.1in}{.1in}}

\setcitestyle{numbers,square}
\begin{document}
\title{Critical Non-Hermitian Edge Modes}
\author{Kunling Zhou}
\thanks{These authors contribute equally to this work.}

\affiliation{School of Physics, Huazhong University of Science and Technology, Wuhan 430074, P. R. China}

\affiliation{Hunan Provincial Key Laboratory of Flexible Electronic Materials Genome Engineering,
School of Physics and Electronic Sciences, Changsha University of Science and Technology, Changsha 410114, P. R. China}

\author{Zihe Yang}
\thanks{These authors contribute equally to this work.}


\affiliation{School of Physics, Huazhong University of Science and Technology, Wuhan 430074, P. R. China}

\author{Bowen Zeng}
\email[]{zengbowen@csust.edu.cn}
\affiliation{Hunan Provincial Key Laboratory of Flexible Electronic Materials Genome Engineering,
School of Physics and Electronic Sciences, Changsha University of Science and Technology, Changsha 410114, P. R. China}

\author{Yong Hu}
\email[]{huyong@hust.edu.cn}
\affiliation{School of Physics, Huazhong University of Science and Technology, Wuhan 430074, P. R. China}

\makeatletter
\newcommand{\rmnum}[1]{\romannumeral #1}
\newcommand{\Rmnum}[1]{\expandafter\@slowromancap\romannumeral #1@}
\makeatother

\begin{abstract}
 We unveil a unique critical phenomenon of topological edge modes in non-Hermitian systems, dubbed the critical non-Hermitian edge modes (CNHEM). Specifically, in the thermodynamic limit, the eigenvectors of edge modes jump discontinuously under infinitesimal on-site staggered perturbations. The CNHEM arises from the competition between the introduced on-site staggered potentials and size-dependent non-reciprocal coupling between edge modes, and are closely connected to the exceptional point (EP).  As the system size increases, the coupling between edge modes decreases while the non-reciprocity is enhanced, causing the
 eigenvectors to gradually collapse toward the EP. However, when the on-site potentials dominate, this weakened coupling assists the eigenvectors to stay away from the EP. Such a critical phenomenon is absent in Hermitian systems, where the coupling between edge modes is reciprocal. \newline 
 \newline
 \textbf{non-Hermitian topological edge modes, critical phenomenon, exceptional point, size-dependent coupling }\newline
 \newline 
 \textbf{PACS number(s):} 68.35.Rh, 71.23.An, 71.15.-m
\end{abstract}

\maketitle

\section{Introduction}
Critical systems, referring to systems at the critical point or lying at the phase boundary, have attracted much attention due to their profound theoretical implications and wide practical applications~\cite{vidal2003entanglement,korepin2004universality,dziarmaga2005dynamics,hertz2018quantum,chang2020entanglement}. 
In quantum critical systems that involve topological phase transitions, the eigenvalues and eigenvectors of the associated Hamiltonian typically undergo continuous changes across the critical point~\cite{hasan2010colloquium,qi2011topological,asboth2016short}. 
Recently, Li~\textit{et al.} reported a novel critical phenomenon in coupled two Hatano-Nelson chains~\cite{hatano1996localization,li2020critical}, called the critical non-Hermitian skin effect (CNHSE), where the eigenvalues and eigenvectors discontinuously jump under 
infinitesimal inter-chain coupling in the thermodynamic limit~\cite{yang2020non,li2020critical}. In a single Hatano-Nelson chain, the non-reciprocal coupling drives the bulk state to accumulate at the boundary, a phenomenon known as the non-Hermitian skin effect~\cite{lee2016anomalous,xiong2018,kunst2018biorthogonal,yao2018edge,longhi2019probing,ashida2020non,borgnia2020non,RevModPhys.93.015005,zhang2022review,ding2022non,zhang2022universal,lin2023topological}. While in coupled two Hatano–Nelson chains with a piling up of states at distinct ends, the inter-chain coupling leads to the gradual disappearance of the NHSE as the system size increases~\cite{qin2023universal,rafi2022critical,kawabata2023entanglement,cai2024non,zhang2025magnetically,liu2024non}. These results suggest that the effective coupling in such systems is dependent on size, offering an perspective for understanding this critical behavior~\cite{li2020critical,yokomizo2021scaling}. As pointed by Yokomizo~\textit{et al.}, the increase of system size amplifies the effective coupling between two chains~\cite{yokomizo2021scaling}, so that an arbitrarily small coupling strength can become effectively large in the thermodynamic limit, ultimately triggering the CNHSE.

Another well-known phenomenon associated with size-dependent coupling---though not necessarily linked to the above critical behavior---is the hybridization of edge modes (EMs) in topological insulators~\cite{hasan2010colloquium,qi2011topological,asboth2016short}. In the one-dimensional Su-Schrieffer–Heeger (SSH) model, the Hermitian coupling between topological EMs decays exponentially with the system size~\cite{su1979solitons,su1980soliton,nakada1996edge,son2006energy,asboth2016short,ding2024robustness}. In the non-Hermitian SSH model, this coupling becomes not only size-dependent but also non-reciprocal~\cite{lee2016anomalous,cheng2022competition,wang2020defective}. Such a coupling induces degenerate eigenvalues and the collapse of eigenvectors for EMs in the thermodynamic limit,  signifying the emergence of exceptional point (EP)~\cite{heiss2012physics,hodaei2017enhanced,miri2019exceptional,RevModPhys.93.015005,ding2022non,zhou2025non}. Remarkably, two EMs become distinct even under an infinitesimal coupling~\cite{wang2020defective}.  These observations raise several important questions. First, could this be a critical phenomenon similar to the CNHSE? Furthermore, how can this size-dependent effect be harnessed to design a critical phenomenon for non-Hermitian EMs? Another key question concerns the phase diagram that characterizes such a critical behavior. Although some studies have reported the existence of a critical size~\cite{li2020critical,yokomizo2021scaling,qin2023universal}, a comprehensive perturbation–size phase diagram is still lacking. 

In this work, we demonstrate a critical phenomenon unique to non-Hermitian EMs, which we refer to as the critical non-Hermitian edge modes (CNHEM). The CNHEM is realized by introducing on-site staggered perturbations into the one-dimensional non-Hermitian SSH model. Analytically, we derive the wavefunction of EMs and their size-dependent coupling. This coupling weakens as the system size increases, whereas its inherent non-reciprocity becomes 
increasingly pronounced. We further construct a perturbation–size phase diagram that characterizes the CNHEM. In the absence of the perturbation, the non-reciprocity becomes unidirectional in the thermodynamic limit, driving the system toward an EP. In contrast, when the staggered perturbation is present and the system size exceeds a critical length, two EMs become gradually separated due to their decreased coupling. Our results offer a comprehensive understanding of the CNHEM and provide a foundation for the control and engineering of non-Hermitian EMs, as well as for exploring related critical phenomena in non-Hermitian systems.

\begin{figure}
    \centering
    \hspace*{0.0cm}
    \includegraphics[width = 0.475 \textwidth]{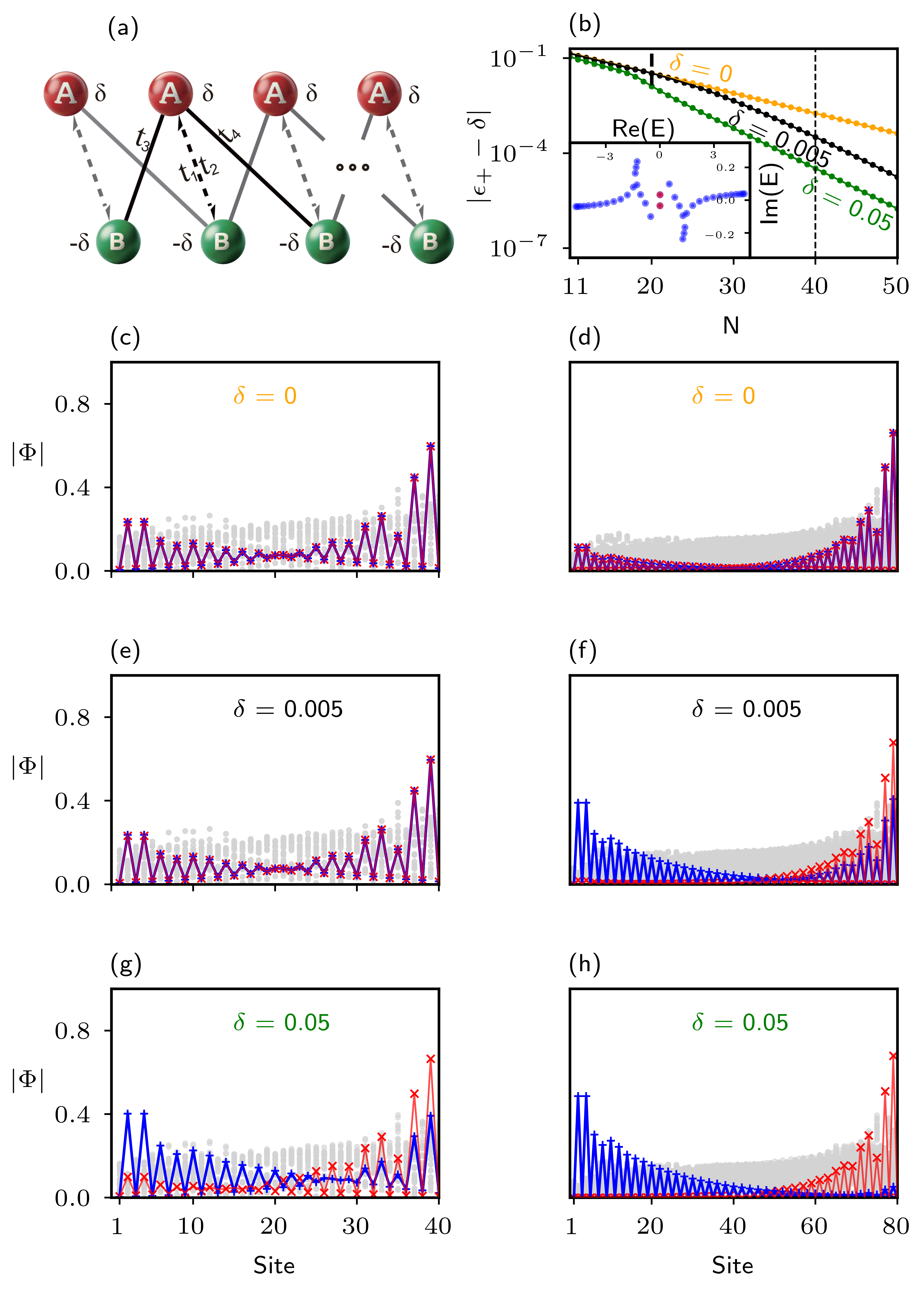}
    \caption{(a) Coupled chains $``A"$ and $``B"$ with Hermitian coupling $t_3,t_4$, non-Hermitian coupling $t_1,t_2$ and staggered on-site potentials $\delta, -\delta$. The parameters chosen are $t_1=2,t_2=1.5,t_3=e^{-\pi i/6},t_4=2$. (b) The dependence of $\left| \epsilon_{+} - \delta \right|$ on size when $\delta = 0$ (yellow points), $\delta = 0.005$ (black points) and $\delta = 0.05$ (green points). The inset shows the spectrum under $\delta=0,N=20$ under open boundary condition with red dots denoting the energy of EMs. To exhibit the CNHEM, the distributions of numerically calculated EMs are illustrated with blue $``+"$ and red $``\times"$ in two system sizes $N = 20, 40$ (site numbers 40, 80), when $\delta = 0$ [(c), (d)], $\delta = 0.005$ [(e), (f)] and $\delta = 0.05$ [(g), (h)], while the wavefunction of bulk states are denoted by gray regions. In (b-h), corresponding to these numerical data points, the analytic results are represented by lines of the same color.}
    \label{fig-model}
\end{figure}

\section{Size-dependent phase transition for the CNHEM}
To illustrate the generality of CNHEM, we employ a widely used model~\cite{yokomizo2021scaling,hou2022deterministic}, as depicted in Fig.~\ref{fig-model}(a). The Hamiltonian of this system is given by
\begin{align}\label{lattH}
    H= \sum_n &t_1a_n^{\dagger}b_n+t_2b_n^{\dagger}a_n+t_3a_{n+1}^{\dagger}b_n+t_3^* b_n^{\dagger}a_{n+1}\nonumber \\+&t_4b_{n+1}^{\dagger}a_n+t_4^*a_n^{\dagger}b_{n+1}+\delta a_n^{\dagger}a_n-\delta b_n^{\dagger} b_n,
\end{align}
where $t_{1}-t_{4}$ are coupling strength with only $t_1 \neq t_2^*$ being the non-Hermitian term and $
(\delta,-\delta)$ are the on-site staggered potentials introduced on $(``A",``B")$ sublattice. In the presence of nontrivial band topology under suitable parameters~\cite{yao2018edge,yin2018geometrical,yokomizo2020non,yang2025inverse}, this system exhibits two EMs with their eigenvalues denoted by $ \epsilon_{\pm}$ symmetrically distributed with respect to the origin point, as shown in the inset of Fig.~\ref{fig-model}(b). When $\delta = 0$, the absolute value of eigenvalues $\left| \epsilon_{\pm} \right|$ decreases exponentially as the size increases. Fig.~\ref{fig-model}(c) shows the numerically calculated eigenvectors of two EMs (blue $``+"$ and red $``\times"$) in the system size $N=20$. These two eigenvectors exhibit similar distributions at each site, which implies a high degree of overlapping. Such highly overlapped eigenvectors remain preserved when the system size increases, such as $N=40$ in Fig.~\ref{fig-model}(d).

Now consider small on-site staggered potentials $\delta = 0.005$, the scaling relation of $\left| \epsilon_{+} - \delta \right| $ in Fig.~\ref{fig-model}(b) exhibits same behavior as that in the $\delta=0$ case for small $N$. Here, the real part of $\epsilon_{+}$ is taken to have the same sign as $\delta$. However, when the system size exceeds a critical length, a distinct scaling relation for the eigenvalues emerges with a steeper slope, implying the occurrence of a phase transition. Although the chiral symmetry is broken by $\delta$, these EMs still exist as shown in Fig.~\ref{fig-model}(e)-(f), which are protected by the global Berry phase~\cite{liang2013topological,yang2025inverse}. Correspondingly, before the phase transition, two EMs remain highly overlapped in the system size $N=20$ [Fig.~\ref{fig-model}(e)]; after the phase transition, by contrast, the eigenvectors of two EMs become separately distributed at two ends when $N=40$ [Fig.~\ref{fig-model}(f)]. A larger $\delta = 0.05$ also leads to a similar phase transition but with a smaller critical size, as shown in Fig.~\ref{fig-model}(b). Consequently, the eigenvectors of two EMs separate from each other when $N=20$ [Fig.~\ref{fig-model}(g)], and a further increase in system size leads to a more pronounced separation between them [Fig.~\ref{fig-model}(h)].

These results imply that a small $\delta$ merely postpones the occurrence of phase transition, as the system size increases. It can be expected that an infinitesimal $\delta$ in the thermodynamic limit can also lead to such a phase transition. Such a critical phenomenon for EMs is dubbed the CNHEM. Though similar to the CNHSE, this CNHEM is independent of the CNHSE, since the behavior of bulk states [gray regions in Figs.~\ref{fig-model}(c-h)] changes little with the increase in the system size.

\section{Analytic solutions for the EMs}
To gain a deep understanding of the CNHEM, we begin by analytically calculating the wavefunctions of the EMs. The eigenstates for model in Fig.~\ref{fig-model}(a) can be constructed as a combination of non-Bloch basis $\beta$~\cite{yao2018edge,yokomizo2020non}, 
\begin{align}\label{eq-basis}
    \Phi(n)=\mqty[\Phi_A(n) \\\Phi_B(n)]=\sum_i \mqty[\phi_{A_i} \\ \phi_{B_i}] \beta_i^n.
\end{align}
Here, the index $i$ represents the number of roots, which is further determined by the characteristic equation later. Then, Eq.~\eqref{lattH} can be written in the non-Bloch form,
\begin{align}
\label{eq-blochform}
    H=\left[\begin{array}{cc} 
\delta & R_{+}(\beta) \\
R_{-}(\beta) & -\delta
\end{array}\right],
\end{align}
with $R_{+}(\beta) = t_1+t_3 /\beta+t_4^*\beta$ and $R_{-}(\beta) = t_2+t_4 /\beta+t_3^*\beta$, and the characteristic equation is obtained as \begin{align}
\label{eq-SSHboundary}
    E^2-\delta^2=R_{+}(\beta)R_{-}(\beta).
\end{align}
For each eigenvalue $E$, there are four $\beta$s, denoted as $\{\beta_1,\beta_2,\beta_3,\beta_4\}$ sorted by their modulus $ |\beta_1| \leq |\beta_2| \leq |\beta_3| \leq |\beta_4|$. Substituting Eq.~\eqref{eq-basis} into the 
boundary equations~\cite{yao2018edge,yokomizo2020non}  yields $\sum_{i=1}^4 \phi_{A_i} = \sum_{i=1}^4 \phi_{B_i} = \sum_{i=1}^4 \phi_{A_i} \beta_i^{N+1} \sum_{i=1}^4 \phi_{B_i} \beta_i^{N+1}=0$.

Utilizing the relationship between $\phi_{A_i}$ and $\phi_{B_i}$, the secular equation can be written as 
$M\phi_A = 0$ with $M=$
\begin{align}
\label{eq-SSHedgeboundary}
    \left|\begin{array}{cccc}
1 & 1 & 1 & 1 \\
\frac{R_{-}(\beta_1)}{E+\delta} & \frac{R_{-}(\beta_2)}{E+\delta} & \frac{E-\delta}{R_{+}(\beta_3)} & \frac{E-\delta}{R_{+}(\beta_4)} \\
\beta_1^{N+1} & \beta_2^{N+1} & \beta_3^{N+1} & \beta_4^{N+1} \\
\frac{R_{-}(\beta_1)\beta_1^{N+1}}{E+\delta}   & \frac{R_{-}(\beta_2)\beta_2^{N+1}}{E+\delta} & \frac{(E-\delta)\beta_3^{N+1}}{R_{+}(\beta_3)} & \frac{(E-\delta)\beta_4^{N+1}}{R_{+}(\beta_4)}
\end{array}\right|
\end{align}
and $\phi_A = \left[\phi_{A_1},\phi_{A_2}, \phi_{A_3},\phi_{A_4}\right]^{T}$.
The solutions exist when det$(M)=0$. 
The bulk solutions in the thermodynamic limit correspond to the condition
$|\beta_2| = |\beta_3|$, the set of which constitutes the generalized Brillouin zone~\cite{yao2018edge,yokomizo2020non,yang2020non,zhang2020correspondence,okuma2020topological,yu2024non,zhou2024abnormal}.

However, there may exist special solutions with det$(M)=0$ and $|\beta_2| < |\beta_3|$, which correspond to the EMs~\cite{yang2025inverse}. In our previous work, we have proven that the condition of the existence of EMs in such a system requires that both two roots for one of $R(\beta)=0$ are larger (smaller) than those of the other~\cite{yang2025inverse}.  Without loss of generality, we assume that the two roots of $R_{+}(\beta)=0$ [$R_{-}(\beta)=0$] correspond to $\{\beta_1,\beta_2\}$ [$\{\beta_3,\beta_4\}$]. 
Note that the above discussions hold only in the thermodynamic limit. In the case of finite size, $\epsilon_{+} - \delta \to 0$ with the increase in the system size, it can be reasonably speculated that 
\begin{align}
    R_{+}(\beta_1),R_{+}(\beta_2),R_{-}(\beta_3),R_{-}(\beta_4) \propto \epsilon_{+}-\delta.
\end{align}

For solving the EMs, we can only focus on two dominated terms containing $(\beta_4\beta_3)^{N+1}$ and $(\beta_4\beta_2)^{N+1}$ in $\det(M)$, which cancel with each other, yielding the eigenvalues of EMs
\begin{align}\label{eq-eigenvalue}
    \epsilon_{\pm} =\pm \sqrt{c^2(\beta_2/\beta_3)^{N+1}+\delta^2},
\end{align}
with
\begin{align}
   \frac{1}{c^2}=(\frac{1}{R_{+}(\beta_4)}-\frac{1}{R_{+}(\beta_3)})(\frac{1}{R_{-}(\beta_1)}-\frac{1}{R_{-}(\beta_2)}).
\end{align}
The analytic results from Eq.~\eqref{eq-eigenvalue} agree with numerical results, as shown in Fig.~\ref{fig-model}(b). When $\delta=0$, the system holds chiral symmetry and these two eigenvalues exhibit a unified scaling relation $c(\beta_2/\beta_3)^{\frac{N+1}{2}}$. When $\delta\neq 0$, the chiral symmetry is broken, and the scaling relation becomes size-dependent. In the small-sized region, $\delta^2 \ll c^2(\beta_2/\beta_3)^{N+1}$, it can be expected that a similar scaling relation for $\left| \epsilon_{+} -\delta \right|$. But as the system size increases, the former term inside the square in Eq.~\eqref{eq-eigenvalue} decreases until $c^2\left(\beta_2/\beta_3 \right)^{N+1} \ll \delta^2$. Beyond this size, the eigenvalues can be approximated as 
\begin{align}\label{eq-edgeenergy}
    \epsilon_{\pm} = \pm\qty[\delta +\frac{c^2}{2\delta }\qty(\frac{\beta_2}{\beta_3})^{N+1}],
\end{align}
leading to a different scaling relation $\epsilon_{+} -\delta \propto (\beta_2/\beta_3)^{N+1}$. 
These results imply that when the chiral symmetry is broken, two totally different asymptotic behaviors occur as the system size increases and the critical size depends on $\delta$.

The phase transition also leads to distinct distributions of EMs, as discussed in the following. We notice that in Eq.~\eqref{eq-SSHedgeboundary} the former (last) two terms  dominate in the second (third) row, thereby $\phi_{A_1}\approx - 
\left(R_{-}(\beta_2)/ R_{-}(\beta_1)\right)\phi_{A_2}$ and $\phi_{A_3} = - \left(\beta_4/\beta_3 \right)^{N+1}\phi_{A_4}$.
Substituting into Eq.~\eqref{eq-SSHedgeboundary}, the first and fourth rows give the same ratio, 
\begin{equation}
    \eta = \frac{\phi_{A_3}}{\phi_{A_2}} \approx \frac{R_{-}(\beta_2)-R_{-}(\beta_1)}{R_{-}(\beta_1)},
\end{equation}
which further verifies the above approximate treatment. Then the wavefunctions of EMs (denoted by $e$) on the site $``A"$ and $``B"$ are 
\begin{equation}
\begin{aligned}
\label{eq-generalcase-edgestae}
    \Phi_{Ae}(n) &= \beta_2^n-\frac{R_{-}(\beta_2)}{R_{-}(\beta_1)}\beta_1^n  + \eta\beta_3^n- \qty(\frac{\beta_3}{\beta_4})^{N+1}  \eta\beta_4^n, \\
    \Phi_{Be}(n) &= \frac{R_{-}(\beta_2)}{\epsilon_{\pm}+\delta} \beta_2^n- \frac{R_{-}(\beta_2)}{\epsilon_{\pm}+\delta}\beta_1^n    \\&+\frac{\epsilon_{\pm}-\delta}{R_{+}(\beta_3)}\eta\beta_3^n-\frac{(\epsilon_{\pm}-\delta)}{R_{+}(\beta_4)}\qty(\frac{\beta_3}{\beta_4})^{N+1} \eta \beta_4^n.
\end{aligned}
\end{equation}
The wavefunctions for two EMs with different energy $\epsilon_{\pm}$ are denoted by $\ket{e_{\pm}}$. These analytic wavefunctions (lines) are in agreement with the numerical results (data points), as shown in Figs.~\ref{fig-model}(c-h). Now we can evaluate the influence of phase transition on the wavefunctions by analyzing the localized direction of the EMs. When $\delta = 0, \epsilon_{\pm} = \pm c(\beta_2/\beta_3)^{\frac{N+1}{2}} $, the EMs located at the left and right ends are 
$\Phi_{Ae}(1) \propto 1$, $\Phi_{Ae}(N) \propto \beta_3^N$, $\Phi_{Be}(1) \propto (\beta_3/\beta_2)^{\frac{N}{2}}$ and $\Phi_{Be}(N) \propto (\beta_2\beta_3)^{\frac{N}{2}}$. Such distributions can be classified into four cases: 
{\small
\begin{itemize}[leftmargin=5mm]
    \item $\abs{\beta_2}\!<\!\abs{\beta_3}\!<\!1$, $\Phi_{Be}(1)\!>\!\Phi_{Ae}(1)\!>\!\Phi_{Ae}(N)\!>\!\Phi_{Be}(N)$;
    \item $\abs{\beta_2\beta_3}\!<\!1\& \abs{\beta_3}\!>\!1 $, $\Phi_{Be}(1)\!>\!\Phi_{Ae}(N)\!>\!\Phi_{Ae}(1)\!>\!\Phi_{Be}(N)$; 
    \item $\abs{\beta_2\beta_3}\!>\!1 \& \abs{\beta_2}\!<\!1$, $\Phi_{Ae}(N)\!>\!\Phi_{Be}(1)\!>\!\Phi_{Be}(N)\!>\!\Phi_{Ae}(1)$; 
    \item $1\!<\!\abs{\beta_2}\!<\!\abs{\beta_3}$, $\Phi_{Ae}(N)\!>\!\Phi_{Be}(N)\!>\!\Phi_{Be}(1)\!>\!\Phi_{Ae}(1)$. 
\end{itemize}}
In summary, the distribution of EMs is determined by $\abs{\beta_2\beta_3}$. When $\abs{\beta_2\beta_3}>1$ ($\abs{\beta_2\beta_3}<1$), the EMs are localized at the right (left) side. In the thermodynamic limit, from an overall perspective of site distribution, the EMs are dominated by $\Phi_{Ae}(n)$ when $\abs{\beta_2\beta_3}<1$ and $\Phi_{Be}(n)$ when $\abs{\beta_2\beta_3}>1$.

These behaviors are significantly altered when $\delta\neq  0$ after the phase transition. Substituting the energy of EM Eq.~\eqref{eq-edgeenergy} into Eq.~\eqref{eq-generalcase-edgestae}, we have 
\begin{equation}\label{eq-poepsilon}
    \begin{aligned}
     \Phi_{Be}(n) &=  \frac{-c^2}{2R_{+}(\beta_3)\delta}\qty(\frac{\beta_2}{\beta_3})^{N+1}\eta\beta_3^n+  \frac{R_{-}(\beta_2)}{2\delta}\beta_2^n \\& - \frac{-c^2}{2R_{+}(\beta_3)\delta}\qty(\frac{\beta_2}{\beta_4})^{N+1} \eta \beta_4^n -\frac{R_{-}(\beta_2)}{2\delta} \beta_1^n,  \\
    \end{aligned}
\end{equation}
for $\epsilon_{+}$. When $\abs{\beta_3}<1$, $\{\Phi_{Be}(1),\Phi_{Ae}(1)\}>\{\Phi_{Be}(N),\Phi_{Ae}(N)\}$, the wavefunctions are skewed to the left end. While when $\abs{\beta_3}>1$, $\Phi_{Ae}(N) \propto \beta_3^N $ at the right end dominates the distribution. 
For $\epsilon_{-}$, 
\begin{equation}\label{eq-mepsilon}
    \begin{aligned}
    \Phi_{Be}(n) &=  \frac{-2\delta}{R_{+}(\beta_3)}\eta\beta_3^n-\frac{-2\delta}{R_{+}(\beta_4)}\qty(\frac{\beta_3}{\beta_4})^{N+1} \eta \beta_4^n  \\&+ \frac{-2\delta R_{-}(\beta_2)\beta_3^{N+1}}{c^2\beta_2^{N+1}} (\beta_2^n-  \beta_1^n).  \\
    \end{aligned}
\end{equation}
Similar analyses for the distribution imply that it is necessary to compare the magnitude between $(\beta_3/\beta_2)^{N+1}$ at the left end and $\beta_3^{N}$ at the right end, and thereby the localized direction of this EM is determined by $\abs{\beta_2}$. Meanwhile, in the thermodynamic limit, the site distribution implies that the wavefunction can be approximated by $\Phi_{Ae}(n)$ for $\epsilon_{+}$ and $\Phi_{Be}(n)$ for $\epsilon_{-}$.

\begin{figure}
    \centering
    \hspace*{0.0cm}
    \includegraphics[width = 0.475 \textwidth]{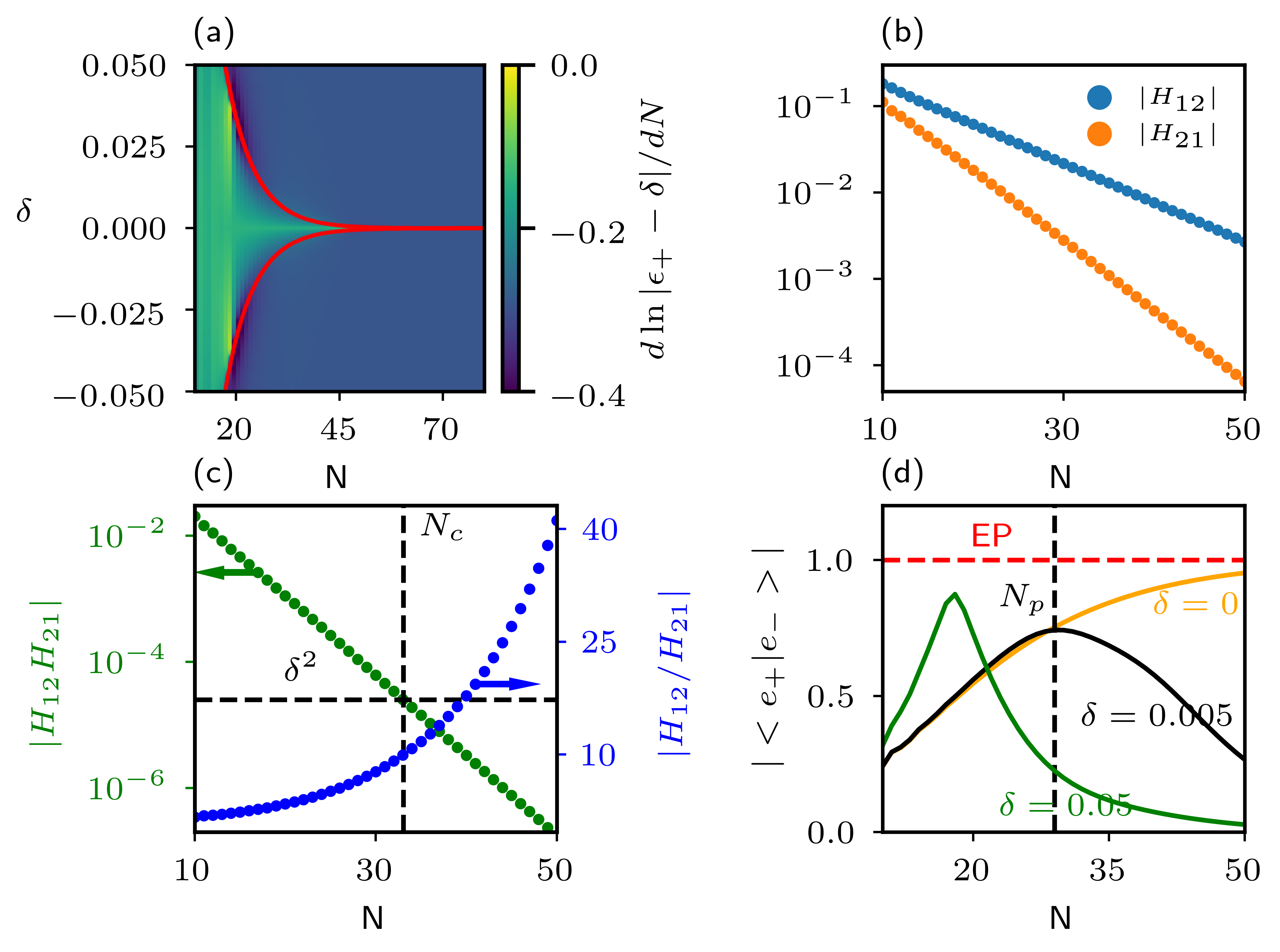}
    \caption{(a) Perturbation-size phase diagram of CNHEM with size-dependent phase boundary. The numerical results (color map) are aligned with the theoretically estimated critical sizes. Under $\delta=0.005$ and the same $t_{1}-t_{4}$ as in Fig.~\ref{fig-model}, the coupling between two EMs in (b) with the increase of system size leads to the reduction of coupling strength $H_{12}H_{21}$ and the enhancement of non-reciprocity $H_{12}/H_{21}$, as shown in (c). (d) The overlapping magnitude between the EMs, where a complete overlapping signifies the emergence of the EP. }
    \label{fig-phase}
\end{figure}
\section{Size-dependent phase boundary and the EP}
Considering different scaling behaviors for the energy of EMs across the critical point, the derivative of $\abs{\epsilon_{+}-\delta}$ with respect to length must have a mutation at the critical size, as shown in Fig.~\ref{fig-phase}. This perturbation-size phase diagram also exhibits a size-dependent boundary, a unique feature for CNHEM. Analytically, the phase transition occurs when $\delta = c(\beta_2/\beta_3)^{\frac{N+1}{2}}$ [see Eq.~\eqref{eq-eigenvalue}]. Consequently, the critical length is estimated as 
\begin{align}\label{criticallength}
    N_c=2\ln (\delta/c) / \ln(\beta_2/\beta_3)-1,
\end{align}
which exactly matches the mutation of the color map, as shown in Fig.~\ref{fig-phase}.

Now we turn to the mechanism of the CNHEM. Both the form of eigenvalues [Eq.~\eqref{eq-eigenvalue}] and eigenvectors [Eq.~\eqref{eq-generalcase-edgestae}] imply that there exists a size-dependent coupling between EMs.  Combined with the analysis of the distributions of the wavefunction [see Eqs.~\eqref{eq-generalcase-edgestae}-\eqref{eq-mepsilon}] in the finite size and thermodynamic limit, this size coupling essentially induces the superposition of $\Phi_{Ae}(n)$ and $\Phi_{Be}(n)$. Consequently, the effective Hamiltonian $H_{e}$ describing the coupling between the EMs can be estimated by investigating the hopping strength between $\Phi_{Ae}(n)$ and $\Phi_{Be}(n)$.

The EMs on the $``A"$ and $``B"$ site can be defined as 
\begin{equation}
    \begin{aligned}
    \ket{A}&=c_A \sum_{n=1}^N \Phi_{Ae}(n) \ket{A,n}, \\
     \ket{B}&=c_B \frac{\epsilon_{\pm} + \delta}{R_{-}(\beta_2)} \sum_{n=1}^N \Phi_{Be}(n) \ket{B,n}.
    \end{aligned}\label{eq:edgestate}
\end{equation}
with  normalized factor 
\begin{equation}
    \frac{1}{\abs{c_{A/B}}^2} \propto \sum_{n=1}^N\abs{\beta_{3/2}}^{2n}  \propto \frac{1-\abs{\beta_{3/2}}^{2N+2}}{1-\abs{\beta_{3/2}}^2}.
\end{equation}
The matrix form of Hamiltonian in Eq.~\eqref{lattH} can also be written as 
\begin{align}\label{eq:H}
    H = \mqty [&\delta I_{N}  &H_{AB} \\ &H_{BA}  &-\delta I_N],
\end{align}
where $H_{AB}$ $(H_{BA})$ represents the coupling from site $``B"$ ($``A"$) to $``A"$ ($``B"$). Their actions on the wavefunction yield
\begin{equation}
    \begin{aligned}
       H_{BA} \ket{A} \simeq &   \frac{c_A R_{-}(\beta_2)}{c_B} \ket{B}, \\
        H_{AB} \ket{B} \simeq &  \frac{c_B(\epsilon_{\pm}^2-\delta^2)}{R_{-}(\beta_2)c_A} \ket{A}.
    \end{aligned}
\end{equation}
It can be seen that $\ket{A}$ and $\ket{B}$ form an invariant subspace for these two EMs. The effective Hamiltonian $H_{e}$ in such a subspace can be estimated by a $2\times2$ matrix,
\begin{align} \label{eq:Hedge}
        H_{e} &\!= \left[\begin{array}{cc}
           H_{11}  & H_{12}  \\
            H_{21} & H_{22} 
        \end{array}\right] =  \left[\begin{array}{cc}
           \bra{A} \delta I_{N} \ket{A} &\bra{A} H_{AB}\ket{B}  \\
           \bra{B} H_{BA}\ket{A} &\bra{B} -\delta I_{N}\ket{B}
        \end{array}\right] \nonumber  \\ 
        & = \left[\begin{array}{cc}
           \delta  &\frac{c_B(\epsilon_{\pm}^2-\delta^2)}{R_{-}(\beta_2)c_A} \\
            \frac{c_A R_{-}(\beta_2)}{c_B}  &-\delta 
        \end{array}\right].
\end{align}
This equation also gives the same eigenvalues as Eq.~\eqref{eq-eigenvalue},  and the normalization eigenvectors 
\begin{align}
    \ket{e_{\pm}}=\frac{\frac{c_B(\epsilon_{\pm}+\delta)}{R_{-}(\beta_2)c_A} \ket{A}+\ket{B}}{\sqrt{\abs{\frac{c_B(\epsilon_{\pm}+\delta)}{R_{-}(\beta_2)c_A}}^2+1}} ,
\end{align} 
which are proportional to Eq.~\eqref{eq-generalcase-edgestae}, confirming our results. The non-reciprocal can be estimated as 
\begin{equation}\label{eq-nonreciprocal}
    \frac{H_{12}}{H_{21}} = \frac{c_B^2 c^2  }{c_A^2R_{-}(\beta_2)^2} \qty(\frac{\beta_2}{\beta_3})^{N+1}, 
\end{equation}
which becomes $H_{12}/H_{21} \propto \left(\beta_2/\beta_3 \right)^{N+1}$ when $\abs{\beta_2}<\abs{\beta_3}<1$, $H_{12}/H_{21} \propto (\beta_2\beta_3)^{N+1}$ when $\abs{\beta_2}<1<\abs{\beta_3}$, and $H_{12}/H_{21} \propto \left(\beta_3/\beta_2 \right)^{N+1}$ when $1<\abs{\beta_2}<\abs{\beta_3}$. Except for some special cases such as $\abs{\beta_2\beta_3}=1$, this non-reciprocity also strongly depends on the size. Meanwhile, the coupling strength between two EMs can be estimated as $H_{12}H_{21} \propto (\beta_2/\beta_3)^{N+1} $. The increase in size reduces the coupling, but enhances the non-reciprocity, as shown in Fig.~\ref{fig-phase}(b) and Fig.~\ref{fig-phase}(c). When $\delta=0$, this coupling leads to gradual overlapping between two EMs close to the EP, as shown in Fig.~\ref{fig-phase}(d). However, when $\delta\neq0$, a similar behavior appears only below the critical length. Beyond the critical length, the on-site terms dominate and the decreased coupling leads to distinct EMs accompanied by a decreased overlapping magnitude $|<e_{+}|e_{-}>|$, as shown in Fig.~\ref{fig-phase}(d).

Let us return to the special cases with $\abs{\beta_2\beta_3}=1$, which actually include the Hermitian counterpart with $t_1=t_2^{*}$ and $\beta_2\beta_3^* = 1$. Even in these cases, a size-dependent phase transition for eigenvalues and eigenvectors is still expected [see Eq.~\eqref{eq-eigenvalue}]. The coupling between the EMs still decreases with the increase of system size and may also be non-reciprocal. However, this non-reciprocity would not turn into unidirectional in the thermodynamic limit [Eq.~\eqref{eq-nonreciprocal}], ensuring that the two EMs remain independent. Therefore, infinitesimal on-site perturbations do not induce abrupt transitions in the eigenvalues and eigenvectors in the thermodynamic limit, which explains the absence of such a critical phenomenon in these scenarios. 

Based on the competition between size-dependent coupling and on-site potentials as discussed above, a good estimation of the trend in Fig.~\ref{fig-phase}(d) can be obtained. Next, we will rigorously evaluate the influence of system size on the overlapping magnitude $|<e_{+}|e_{-}>|$ between edge modes.

\begin{figure}
    \centering
    \hspace*{0.0cm}
    \includegraphics[width = 0.475 \textwidth]{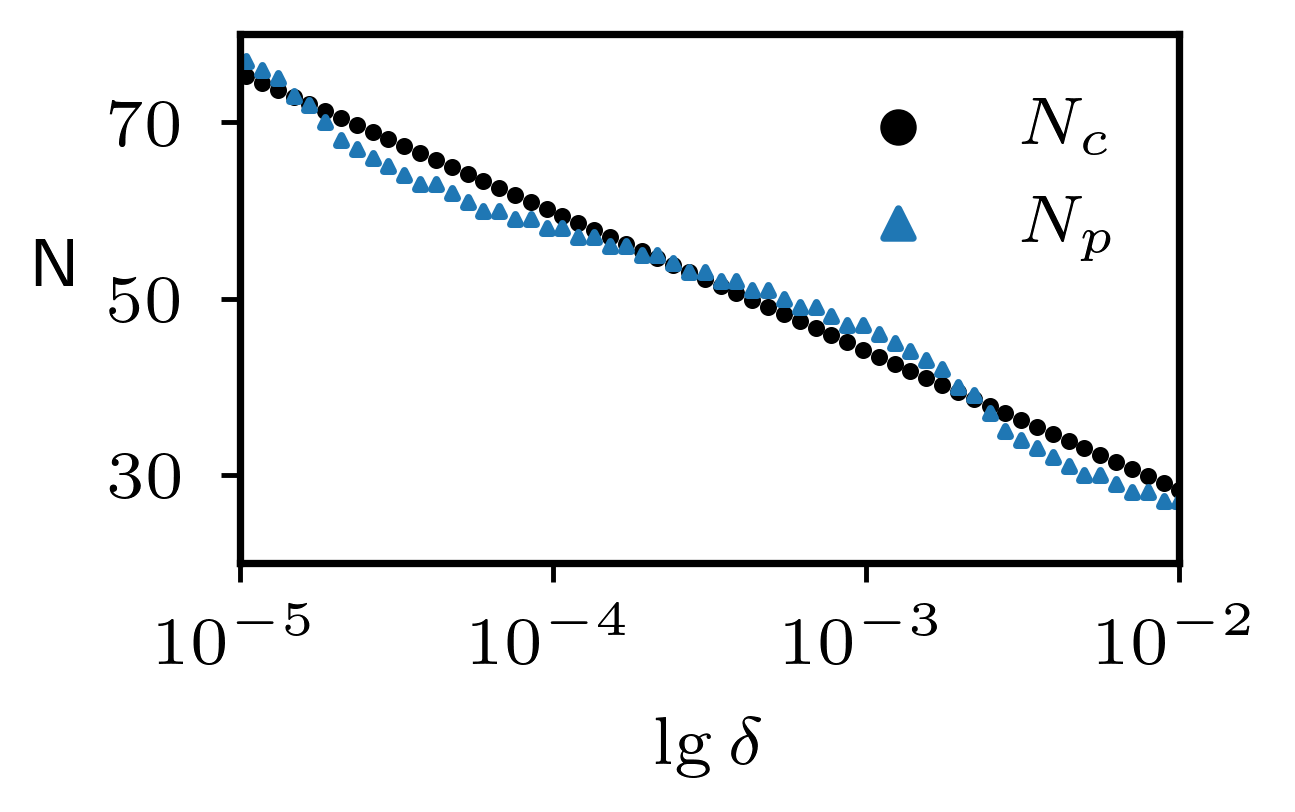}
    \caption{Comparison of critical length $N_c$ (black dots) and peak location $N_p$ (blue triangles) as the function of $\delta$.}
    \label{fig-peak}
\end{figure}

The squared overlapping is given by
\begin{align}
    \abs{\bra{e_+}\ket{e_-}}^2 =\frac{r^2\abs{(\epsilon_{+}^2-\delta^2)}^2+1+2r(\abs{\delta}^2-\abs{\epsilon_{+}}^2)}{r^2\abs{(\epsilon_{+}^2-\delta^2)}^2+1+2r(\abs{\delta}^2+\abs{\epsilon_{+}}^2)},
\end{align}
where $r=\abs{c_B^2/(R_{-}^2(\beta_2)c_A^2)}$. 
The parameters chosen in Fig.~\ref{fig-phase} satisfy $\abs{\beta_3}>1$, $\abs{\beta_2}<1$ and $\abs{\beta_3 \beta_2}>1$. For the case of large $N$, since $r \propto \beta_3^{2N+2}$, we can express $r$ explicitly as $r=r_0\beta_3^{2N+2}$, with $r_0$ being the proportionality factor. Then
\begin{align}
    &\abs{\bra{e_+}\ket{e_-}}^2 \notag \\
    = &1-\frac{\abs{4r_0 \beta_3^{2N+2}\epsilon_{+}^2}}{\abs{c^4r_0^2(\beta_3\beta_2)^{2N+2}}+1+\abs{r_0 \beta_3^{2N+2}}(\abs{\delta}^2+\abs{\epsilon_{+}}^2)} .
\end{align}
When $N \ll N_c$, $\epsilon_{\pm}^2 \simeq c^2(\beta_2/\beta_3)^{N+1} \gg \delta^2$
\begin{align}
\abs{\bra{e_+}\ket{e_-}}^2 
    \simeq  1-\frac{4}{\abs{c^2r_0(\beta_3\beta_2)^{N+1}}} .
\end{align}
As $N$ increases, the overlapping magnitude monotonically increases, approaching the upper bound of 1, as shown in Fig.~\ref{fig-phase}(d). However, when $N \gg N_c$, the energies of edge modes are dominated by $\delta$, thereby 
\begin{align}
\abs{\bra{e_+}\ket{e_-}}^2 \simeq \frac{\abs{r_0^2c^4(\beta_3\beta_2)^{2N+2}}}{\abs{4r_0\beta_3^{2N+2}\delta^2}}\simeq \frac{\abs{r_0c^4\beta_2^{2N+2}}}{\abs{4\delta^2}}
\end{align}
In this regime, the overlapping magnitude decreases with the increase of $N$,  approaching zero in the limit of an infinitely long system. Consequently, for non-zero $\delta$,
there always will be a peak of overlapping between two edge modes, denoted by $N_p$ in Fig.~\ref{fig-phase}(d). 

According to the above analysis, the location of $N_p$ can be estimated by the critical length $N_c$. While $N_p$ is not necessarily equal to $N_c$, since $N_c$  approximates the length scale at which the phase transition occurs. Indeed, $N_c$ and $N_p$ are very close to each other under different $\delta$, as shown in Fig.~\ref{fig-peak}. According to Eq.~\eqref{criticallength}, the critical length is linear with respect to $\lg{\delta}$. While the values of $N_p$ appear to exhibit a small oscillatory behavior around $N_c$.

\section{Conclusions and discussions}
In conclusion, we unveil a critical phenomenon of topological EMs, where arbitrarily small on-site staggered perturbations can induce discontinuous changes in the distribution of EMs in the thermodynamic limit. In finite-sized systems, the phase transition induced by the on-site terms depends on the system size. We analytically solve the wavefunction of EMs and construct the perturbation-size phase diagram, highlighting the size-dependent phase boundary. These size-dependent behaviors stem from the size-dependent coupling between the EMs and their competition with the on-site terms. The size dependence of the coupling manifests in two key aspects: it decreases as the system size increases, but the non-reciprocity increases. This coupling alone leads to the formation of defective EM at the EP, while the introduced on-site potentials in the thermodynamic limit induce mutually independent EMs. Our results not only theoretically demonstrate that the CNHEM is unique to non-Hermitian systems but also provide a perspective for exploring further critical phenomena.

Both the CNHEM and CNHSE exhibit similarities in the discontinuous jump of eigenvectors and size-dependent coupling. The former is dominated by the coupling between edge modes, while the latter is governed by the coupling between skin modes~\cite{li2020critical}. In terms of their mathematical structure, the emergence of both CNHEM and CNHSE can be captured via a $2 \times 2 $ matrix; however, the perturbation terms reside in the diagonal elements for the former, as opposed to the off-diagonal elements for the latter~\cite{li2020critical}.

In essence, experimental platforms, such as active mechanical lattices~\cite{wang2022extended,wang2022non}, phononic or acoustic crystals~\cite{zhang2019non,pu2025non}, and piezophononic media~\cite{gao2020anomalous}, which enable observation and modulation of non-Hermitian edge modes, hold the potential to probe the critical non-Hermitian edge modes. A key observable indicator for such a critical phenomenon is that the phase transition of edge modes depends on both the system size and on-site perturbations, which corresponds to the perturbation-size phase diagram in our work.

To the best of our knowledge, non-Hermiticity affects the properties of edge modes in at least the following three aspects. First, the non-Hermitian skin effect modifies the integration range of the Berry phase and the calculation of topological invariants~\cite{yao2018edge}, leading to the reestablishment of bulk-boundary correspondence. Second, the localization caused by the non-Hermitian skin effect and that induced by topology compete with each other, giving rise to diverse distributions of edge modes~\cite{yang2025inverse}. Finally,  we here demonstrate that the critical phenomenon of edge modes is closely connected to the EP, another unique feature to non-Hermitian systems.

\bigskip
\noindent\textit{This work is supported by the
Natural Science Foundation of Hunan Province (Grant No. 2024JJ6011) and the Quantum Science and Technology-National Science and Technology Major Project (Grant No. 2021ZD0302300).}

\bigskip
\noindent \textbf{Conflict of Interest} The authors declare that they have no conflict of interest.

\bibliography{manuscript.bib}

\end{document}